# Morphology and Photoluminescence of CH$_3$NH$_3$PbI$_3$ Deposits on Nonplanar, Strongly Curved Substrates

*Konstantins Mantulnikovs,† Anastasiia Glushkova,† Peter Matus,† Luka Ćirić,† Marton Kollar,† Laszlo Forro,† Endre Horvath,† and Andrzej Sienkiewicz*,†,‡*

† Laboratory of Physics of Complex Matter, École Polytechnique Fedérale de Lausanne, CH-1015 Lausanne, Switzerlandˇ

‡ ADSresonances SARL, Route de Geneve 60B, CH-1028 Preverenges, Switzerland

Corresponding Author:

*E-mail: andrzej.sienkiewicz@epfl.ch

**ABSTRACT**

Organic–inorganic metal halide perovskites have recently attracted increasing attention as highly efficient light harvesting materials for photovoltaic applications. The solution processability of these materials is one of their major advantages on the route toward fabrication of low-cost solar cells and optoelectronic devices. However, the precise control of crystallization and morphology of organometallic perovskites deposited from solutions, considered crucial for enhancing the final photovoltaic performance, still remains challenging. In this context, here, we report on growing microcrystalline deposits of methylammonium lead triiodide perovskite, CH3NH3PbI3 (MAPbI3), by one-step solution casting on cylinder-shaped quartz substrates (rods) having diameters in the range of 80 to 1800 μm. We show that the substrate curvature has a strong influence on morphology of the obtained polycrystalline deposits of MAPbI3. Specifically, a marked size reduction of MAPbI3 microcrystallites concomitant with an increased crystal packing density was observed with increasing the substrate curvatures. In contrast, although the crystallite width and length markedly decreased for substrates with higher curvatures, the photoluminescence (PL) spectral peak positions did not significantly evolve for MAPbI3 deposits on substrates with different diameters. The crystallite size reduction and a denser coverage of microcrystalline MAPbI3 deposits on cylinder-shaped substrates with higher curvatures were attributed to two major contributions, both related to the annealing step of the MAPbI3 deposits. In particular, the diameter-dependent variability of the heat capacities and the substrate curvature-enhanced solvent evaporation rate seemed to contribute the most to the crystallization process and the resulting morphology changes of MAPbI3 deposits on cylinder-shaped quartz substrates with various diameters. The longitudinal geometry of cylinder-shaped substrates provided also a facile solution for checking the PL response of the deposits of MAPbI3 exposed to the flow of various gaseous media, such as oxygen (O2), nitrogen (N2), and argon (Ar). Specifically, under excitation with λexc = 546 nm, the rapid and pronounced decreases and increases of PL signals were observed under intermittent subsequent exposures to O2 and N2, respectively. Overall, the approach reported herein inspires novel, cylinder-shaped geometries of MAPbI3 deposits, which can find applications in low-cost photo-optical devices, including gas sensors.

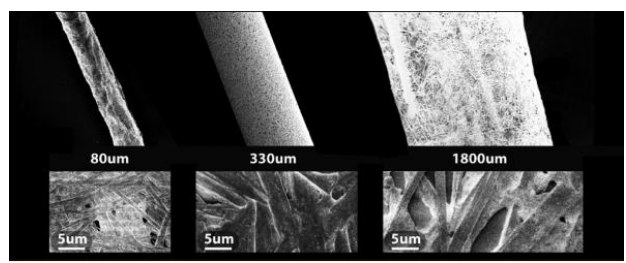

**INTRODUCTION**

The organic−inorganic methylammonium lead triiodide, CH3NH3PbI3 (MAPbI3), is a photovoltaic perovskite material with outstanding optoelectronic properties, such as a very high visible light absorption coefficient, balanced electron and hole transport accompanied by high charge carrier mobilities and long diffusion lengths, and a very strong photoluminescence (PL).1 Not surprisingly, the materials with such favorable optical and electrical properties have been swiftly utilized in solar cells, whose power conversion efficiency rapidly increased from the initial 3.8%2 and 9.7%3 to the recently achieved value of 22.1%,4 thus opening a new avenue for emerging applications of metal−organic perovskites in photovoltaic technology.

Recently, in the context of designing new generations of efficient cost-effective solar cells, significant effort has been devoted toward obtaining uniform and compact thin layers of MAPbI3. In particular, there have been numerous reports on crystallinity and morphology, as well as PL properties, of MAPbI3 obtained either in the form of single-crystalline particles5,6 or polycrystalline films deposited on planar substrates.7,8 It has been generally accepted that the final photovoltaic properties of MAPbI3-based devices are strongly dependent on the film fabrication process, which defines the uniformity and compactness of thin layers of MAPbI3. 9,10

Various fabrication methods of MAPbI3 thin films have been reported to date.11,12 In general, the most commonly used deposition techniques can be categorized into the one-step precursor deposition methods13 and two-step deposition methods.14 It is customarily accepted that the two-step deposition provides better control of the composition, the thickness, and the morphology of the MAPbI3-based films, thus leading to higher photovoltaic performances of the final devices.15 On the other hand, one-step solution-casting protocols seem to be more suitable for manufacturing lowcost MAPbI3-based devices at large scale.16 It has also been widely recognized that obtaining densely packed deposits of MAPbI3 with small grain sizes in the range from hundreds of nanometers to micrometers can be beneficial for numerous prospective optoelectronic applications, including photo-optical gas sensors.17

Although monolithic layers of organometallic perovskites are mostly used in photovoltaic applications, coatings having other morphological characteristics are also being intensively explored. In particular, high power conversion efficiency (PCE) of a coating formed by low-dimensional (1D) MAPbI3 nano- and microstructures has recently been reported by Im et al.18 Specifically, thin layers consisting of nanowires having a 100 nm diameter and a length of several μm were obtained by two-step spin-coating technology.

The two-step deposition route combining the solution process and vapor phase conversion was used for growing polycrystalline MAPbI3 microwires with a length of up to 80 μm that revealed both an excellent light-guiding performance with very low optical loss and band-gap tunability.19 The one-step solution-based synthesis of highly photoactive MAPbI3 nanowires with mean diameters of 50 to 400 nm and lengths up to 10 μm has also been reported by Horvath et al.20

In this context, here, we report on the fabrication and characterization of polycrystalline deposits of MAPbI3 on nonplanar, strongly curved substrates. In particular, with using one-step solution casting21 on curved cylinder-shaped quartz substrates with diameters in the range from 80 to 1800 μm we manufactured the MAPbI3 deposits consisting of polycrystalline microwires with average sizes of 100 nm to 8 μm in diameter and up to ∼150 μm in length.

The optical microscopy and scanning electron microscopy (SEM) imaging of the obtained deposits revealed a pronounced dependence of their morphology on the substrate curvature. Notably, a markedly enhanced surface grain packing, concomitant with smaller average microcrystallite sizes, was observed while depositing MAPbI3 on cylinder-shaped quartz substrates with the smallest diameters (i.e., of 80 and 330 μm). Interestingly, however, the microcrystalline MAPbI3 deposits conserved roughly the same thickness of 12–15 μm throughout the whole range of variability of the substrate diameters.

Since the cylinder-shaped geometry of MAPbI3 deposits facilitates the design of sealed housings, we also explored a possibility of implementing the obtained microcrystalline densely packed MAPbI3 films as prospective photo-optical gas sensors. To this end, a cylindrical substrate having a diameter of 330 μm was covered with MAPbI3 and encapsulated in a section of a slightly larger quartz capillary (3.0 mm i.d./4.0 mm o.d.). This approach provided an easy way for a direct exposure of the MAPbI3-covered substrate to gaseous media, such as $O_2$, $N_2$, and Ar and made it possible to detect gas-dependent changes in the characteristic PL signal of MAPbI3. Overall, this construction revealed its usefulness for photo-optical sensing of gaseous media and may serve as a model of uncomplicated MAPbI3-based gas sensors.

**RESULTS AND DISCUSSION**

**Morphology and Photoluminescence**. Optical microscopy and scanning electron microscopy were used as essential analytical tools for checking the morphology and crystallinity of MAPbI3 deposits on cylinder-shaped quartz substrates. The example optical microscopy images of MAPbI3 deposits acquired for two substrates with diameters of 330 and 1800 μm are shown in Figure 1 (a and b). As can be seen, the morphology of the polycrystalline deposit on a cylindrical substrate having a diameter (DIA) of 330 μm is distinctively different from the one deposited on the 1800 μm DIA substrate. In particular, the MAPbI3 deposit on the 330 μm DIA substrate consists of densely packed micrometer-sized wires, having average cross-sectional diameters of 1.0 μm and ~20 to 30 μm in length. In contrast, the thin film of MAPbI3 deposited on the 1800 μm DIA substrate consists of definitely larger, elongated, and branched fibrous structures, having average dimensions of ~5 to 8 μm in cross-section and ~50 to 70 μm in length.

The luminescence microscopy images, acquired under an excitation of $\lambda_{exc}$ = 546 nm and shown in Figure 1(c and d) also point to the formation of a MAPbI3 deposit having markedly different morphologies for 330 and 1800 μm DIA substrates. In particular, the strongly emitting bright spots correspond to the emission occurring essentially from the terminations of micrometer-sized wires, which are mostly longitudinally aligned on the surface of the 330 μm DIA substrate (Figure 1(c)). In contrast, for the 1800 μm DIA substrate (Figure 1(d)), the PL emission can also be seen from the numerous branch-crossing points, which facilitate the light emission from the otherwise elongated light-channeling structures.[19] It should be noted that the CCD camera used in our experimental setup for luminescence imaging was only able to detect the shortwavelength tail of the MAPbI3 emission. In contrast, the simultaneously acquired PL spectra provided the whole spectral information.

In particular, the PL spectra shown in the insets to Figure 1(c and d), collected for samples deposited on cylinder-shaped substrates with diameters of 330 and 1800 μm and peaking around ~760–770 nm, are characteristic for polycrystalline thin films of MAPbI3.[22] The peak position at 770 nm of the PL spectrum acquired for the deposit coated onto the 1800 μm substrate is slightly red-shifted as reference to the peak position at 761 nm collected for the deposit coated onto the 330 μm substrate. Similarly red-shifted PL spectra have been often reported for larger sizes of MAPbI3 crystallites[23,24] and associated with enhanced self-absorption in the larger crystals.[25] The PL emission peaks shown in the insets to Figure 1(c and d) are also spectrally broadened, and both reveal a shoulder at longer wavelength. Such broadening of the PL spectra and the presence of a long-wavelength shoulder in the PL emission peaks can be ascribed to the presence of the disordered crystalline phases and shallow trapping levels on the grain boundaries.[26,27]

Similar PL spectra were observed in this study for all MAPbI3 deposits on cylinder-shaped substrates (see, for example, Figure S3 for the PL spectrum of MAPbI3 deposited on the 80 μm DIA capillary). Thus, although the crystallite width and length markedly decreased for substrates with higher curvatures, the PL spectral peak positions and the spectral shapes did not significantly evolve for deposits on substrates with different diameters.

To gain further insight into the morphological differences in crystallinity of MAPbI3 layers deposited on cylinder-shaped quartz substrates having various diameters, we performed SEM imaging. The conventional SEM images acquired for MAPbI3 films for the entire range of the substrate diameters (i.e., 80, 330, 400, 700, 1100, and 1800 μm) are shown in Figure 2. As can be seen, the SEM images revealed a marked dependence of the crystallinity and morphology of the thus obtained MAPbI3 deposits on the substrate diameter. In particular, for quartz rods with smallest diameters (80–330 μm), the MAPbI3 thin films consisted of densely packed micrometer wires, whereas for the substrates with larger diameters (400–1800 μm) MAPbI3 formed less densely packed polycrystalline branched and interconnected micrometer-sized ribbon structures, which contained also a large number of voids. Interestingly, however, the microcrystalline MAPbI3 films deposited on cylindershaped quartz substrates conserved roughly the same thickness of 12–15 μm throughout the whole range of variability of the substrate diameters. Thus, in addition to elongated crystallites, which represent the major morphological feature of MAPbI3 deposits in this study, we also observe other forms of polycrystalline MAPbI3, including small spheres and star-like, leaf-like, or dandelion-type structures. In particular, star-like structures systematically occur for MAPbI3 deposits coated onto substrates with larger diameters (lower substrate curvatures). For example, such branched structures can be seen in Figure S9 for MAPbI3 deposits on cylinder-shaped quartz rods having diameters of 400 and 700 μm. In fact, the star-like structures present on cylindrical substrates with larger diameters are similar to these reported by Nie et al. for hot-casting of thins films of MAPbI3 on flat substrates[24] or by Xu et al. for MAPbI3 thin films deposited onto a flat glass substrate (indium tin oxide) covered with a compact layer of TiO2.[28] Moreover, the optical microscopy image in Figure S4(a) also shows the presence of branched and interconnected ribbons of MAPbI3 deposited onto a flat glass surface, while employing the one-step solution coating and analogous annealing conditions as used for coating the cylindrical substrates. It has to be stressed that the "voids", which occur for practically all the deposits presented in Figure 2, are not really free of MAPbI3. As suggested by Xu et al., the space in between the branched and interconnected micrometersized ribbons is also partially filled with smaller, more disordered "spherulitic" crystalline structures of MAPbI3.[28]

Overall, the SEM images presented in Figure 2 revealed a strong relationship between the morphology of MAPbI3 deposits and the substrate curvature. In particular, for the cylinder-shaped quartz substrates with diameters within the range from 80 to 1800 μm, the average sizes of polycrystalline microwires evolved from 100 nm to 8 μm (width) and 5 μm to 150 μm (length).

The mechanisms behind the morphological changes in crystallinity of MAPbI3 deposits as a function of the substrate curvature are complex and involve quite a number of contributions, such as variability of the heat capacities and substrate surface roughness of quartz rods, surface wetting properties,[29] substrate curvature dependent evaporation rates of the solvent (dimethylformamide, DMF), and the local changes in the air flow during annealing of the deposits.

To determine the potential variability of the surface roughness, we used two high-resolution imaging techniques, i.e., SEM and atomic force microscopy (AFM). The SEM images acquired for the uncoated cylindrical quartz substrates of different diameters revealed statistically similar densities of surface defects. Concomitantly, the AFM measurements pointed to a rather low and similar surface roughness (<1.0 nm) for all the quartz substrates. The results of the corresponding SEM and AFM measurements are shown in Figure S13 and Figure S14.

Since in our experimental setup the cylindrical substrates are not in direct contact with the heat source during the annealing process (Figure S1), we also investigated the influence of the variability of substrate heat capacities on the crystallization of MAPbI3 deposits. A rough estimate of the difference in the heat capacities for the substrate diameter range used in this study (80–1800 μm) yields a relatively large factor of 500. Therefore, to gain further insight into the contribution of the variability of substrate heat capacities, the annealing process at 110 °C was performed for both nonpreheated and preheated substrates. To this end, two cylindrical quartz rods with diameters of 80 and 400 μm were simultaneously wetted with the same volumes (∼2.5 μL) of the stoichiometric solution of MAPbI3 precursors in DMF either outside of the hot plate (nonpreheated substrates) or when the rods were already positioned on the hot plate (preheated substrates). The example results of annealing of the MAPbI3 deposits with using these two different scenarios can be viewed in SI Video_Clip_1 and SI Video_Clip_2 for nonpreheated and preheated substrates, respectively.

These video clips reveal that for both scenarios of annealing the crystallization process was definitely much faster for thinner substrates. In particular, for the nonpreheated substrates, the estimated times of MAPbI3 crystallization were 1 s 120 ms and 12 s 520 ms for quartz rod diameters of 80 and 400 μm, respectively, whereas for the preheated substrates the corresponding time lapses were 880 ms and 5 s. Thus, the corresponding crystallization time ratios were ∼12 and ∼5 for the nonpreheated and preheated substrates, respectively. Interestingly, a rough calculation of the difference in the heat capacities for the two quartz substrates with diameters of 80 and 400 μm suggests a factor of ∼25.

It should also be noted that in our experimental setup for annealing the coated quartz rods are positioned at a small distance (∼3.0 mm) from the hot plate (Figure S1). Therefore, concomitantly with a rather large diameter of the hot plate itself (∼25.0 cm), the herein used annealing setup provided comparable air-flow conditions for the quartz cylindrical substrates having different diameters.

We additionally checked the influence of the increasing substrate curvature on the solvent evaporation rate during the thermal annealing of the solution-processed MAPbI3 deposits. The actual influence of the substrate curvature on liquid evaporation rates had been investigated theoretically by William Thomson (also known as Lord Kelvin), who derived the relevant formulas. In principle, the most general form of the Kelvin law (eq 1) is dependent only upon thermodynamic principles and does involve specific properties of materials. It describes the change in vapor pressure due to a curved liquid– vapor interface, such as the surface of a droplet:

$$\frac{RT}{M}\ln\left(\frac{P}{P_0}\right) = \pm 2\frac{\sigma}{d}\left(\frac{1}{r} - \frac{1}{r_0}\right)$$

where R is the gas constant; T, the absolute temperature; M, the molecular weight; σ, the surface energy; d, the density of the liquid; P, the escaping tendency of the substance from a curved surface with radius r; and P0, the escaping tendency from a surface with radius r0, where r0 may be infinitely large, namely, in a flat surface. Thus, as suggested by the Kelvin equation (eq 1), under the same experimental conditions for the thermal annealing, the actual evaporation rate of the solvent from strongly curved convex substrates (e.g., quartz rods with very small diameters) should be higher than from substrates with smaller curvatures.30,31 In fact, similar phenomena to these observed herein have been reported for numerous systems where the curved geometry of the substrates influenced the solvent evaporation rate at both micro- and macroscales.32,33

To further analyze the size variability of polycrystalline MAPbI3 microwires on substrates with different radii (r), while based solely on Kelvin's law, the following assumptions can be made:

1. As evidenced by eq 1, the solvent evaporation rate is proportional to ∼exp(1/r).
2. The nucleation rate of MAPbI3 crystallites should correlate with the solvent evaporation rate.

These assumptions suggest that the average sizes of the formed microwires should also correlate, at least partially, with the substrate radii. Therefore, one should expect crystallites with smaller sizes for the cylindrical substrates with the decreasing substrate radii (i.e., the increasing substrate curvatures, 1/2r). Thus, the average sizes of MAPbI3 microwires should correlate with ∼exp(−1/r). Based on these assumptions, the estimated vapor pressure change, while varying the substrate diameters in the range of 80 to 1800 μm, is on the order of ∼2.5%.

The experimentally measured dependencies of average characteristic sizes (widths and lengths) of MAPbI3 microwires on the substrate curvature are plotted in Figure 3. The corresponding histograms of the crystallite size distributions for each substrate diameter can be found in the Supporting Information (Figures S10 and S11). Thus, the exponential decays shown in Figure 3 point to a strong correlation of the characteristic crystallite sizes of MAPbI3 microwires with the substrate curvature.

Overall, these findings have implications for the field of solution-processed deposition of microcrystalline organometallic perovskites, because they demonstrate that casting and crystallization processes conducted under the same reaction conditions, with the same reaction time, do not produce equivalent deposits; rather, they are dependent on the substrate geometry.

**Influence of Gas Flow on Photoluminescence of MAPbI3 Microwires**. The dense coverage with micrometersized crystallites, concomitant with a large available active surface, obtained on quartz rods with the smallest diameters (80 and 330 μm), encouraged us to check the utility of the cylinder-shaped geometry of MAPbI3 deposits for prospective detection of gaseous media. Specifically, for designing a model photoluminescent gas-sensing device, a MAPbI3-covered quartz rod with a 330 μm diameter was chosen and centrally positioned in a short section of a larger quartz capillary (3.0 mm i.d./4 mm o.d.), which served as an external, lighttransparent sample holder (Figure 4(a)). For further technical details see Figure S2.

This arrangement enabled us to expose the deposits of micrometer-sized MAPbI3 wires to various gaseous media and simultaneously collect their PL spectra. The example timeevolution of the photoluminescent response of the freshly prepared MAPbI3 deposit on a 330 μm diameter quartz rod under continuous illumination with green light (λexc = 546 nm) and intermittent exposure to O2 and N2 is shown in Figure 4(b).

As can be seen in Figure 4(b), while interchanging successively the gas flow from N2 to O2, the photoluminescent response of the MAPbI3 deposit markedly changes. Specifically, under intermittent exposures to N2 and O2, the corresponding PL signals undergo a pronounced intensity increase or decrease, respectively. For easier comparison, the overlapped plots of the decreasing and increasing portions of the photoluminescent responses under the intermittent flows of O2 and N2 are shown in Figure 4(c and d), respectively. The partial decay and the subsequent partial recovery of the PL signal under exposure to the flow of O2 and N2, respectively, was reproducible and could be observed several times during the experiment.

Even though the times of exposure to the two abovementioned gaseous media were not the same, fitting the exponential decaying (rising) function allowed us to extract the characteristic times corresponding to a decrease (increase) of the PL signal amplitude by ca. 2.7 times. It is then evident that the subsequent exposures to O2 result in rapid decays of the PL signals, which occur on a time scale of ~20 s (Figure 4(c)). In contrast, after switching back to N2, the photoluminescent response undergoes a phase of recovery, i.e., restores the initial PL amplitude as in the previous cycle. This process is definitely slower and occurs on a time scale of ~130 s (Figure 4(d)). It is worth pointing out that the plot with the steepest slope (red trace in Figure 4(d)) and the corresponding rise time of ~45 s was recorded during the initial light-soaking treatment under exposure to the flow of N2. The relative changes in the PL signal intensity shown in Figure 4(b) were found to be 50–60%.

Some of the previously reported gas sensors have shown weaker performance in terms of response times as well as usually required a more sophisticated engineering. For example Addabbo et al.[34] reported on a YCoO3-based sensor of CO, NO, and NO2 via measurements of resistance. In their device they used a heated sensing material: the corresponding working temperatures were 180 °C for sensing NOx and more than 200 °C for CO. These devices at their best have response times of 0.5–1.5 min and recovery times of 0.6–4.7 min. Another study on gas sensing, this time using the perovskite material MAPbBr3, has been reported by Fang et al.[35] Their device was based on the intensity of the PL signal and was sensitive to air, dry O2, and moist N2. The response times reported for this device were on the order of 1000 s or longer. Compared to the above-mentioned reports, the response times reported herein are definitely shorter, being in the same range as those reported for more sophisticated devices having, for example, a heated substrate.[36]

The phenomenon of the partial decay of the PL signal under exposure to O2 has been often reported in the context of the studies of the influence of light and oxygen on the stability of MAPbI3-based photoactive layers. Although there is a consensus that, in combination with light, oxygen molecules mainly contribute to quenching of the PL emission in microcrystalline deposits of MAPbI3,[37] the proposed plausible mechanisms explain only partially the corresponding phenomena.[38,39] Moreover, there is also growing experimental evidence that in the context of the monocrystalline MAPbI3 an opposite effect, i.e., the PL enhancement, upon exposure to O2 can also be observed.[40,41]

The rapid diffusion of oxygen into the microcrystalline deposits of MAPbI3 in combination with photoexcited carriers is considered to be the dominant factors resulting in an enhancement of trapping sites responsible for nonradiative charge recombination.[42,43] In particular, the total diffusion times into thin layers (typically 500 nm thick) of MAPbI3 have been found to occur on a time scale of 5–10 min.[43]

Moreover, as can also be seen in Figure 4(b), the subsequent cycles of gas switching from N2 to O2 result in an overall decrease of the photoluminescent response observed during the recovery phase (exposure to N2), as well as in an overall increase of the PL signals acquired at the end of the rapid decay phase (exposure to O2). This might suggest that although the observed changes in the photoluminescent responses upon successive exposures to O2 are largely reversible, there is a longterm oxygen-induced photodegradation of MAPbI3.

Thus, at least two different scenarios of oxygen interactions with microcrystalline deposits of MAPbI3 can be invoked. First, a rapid penetration and diffusion of O2 in parallel with continuous illumination with above-band-gap photons (hν > 1.6 eV) can activate deep trap centers of the photoexcited carriers, thus leading to an efficient and reversible PL quenching. Second, the diffusion of oxygen molecules into polycrystals of MAPbI3 and their interaction with the photoexcited carriers (electrons) may result in formation of superoxide radical species (O2 •−), which are critical to both the oxygen-induced degradation of the host lattice and concomitant irreversible decrease of the PL emission of MAPbI3. 43

**CONCLUSION**

In brief, this work demonstrates for the first time the marked influence of the substrate curvature on the crystallinity and morphology of microcrystalline deposits of MAPbI3 prepared via one-step solution casting.

In particular, MAPbI3 layers deposited on cylinder-shaped quartz substrates with small diameters (80 and 330 μm) proved to be densely packed and consisted of micrometer-sized, needle-shaped wires (1 μm thick and 20 μm long). In contrast, for cylinder-shaped quartz substrates with larger diameters (from 400 to 1800 μm), the variability of the crystalline layers of MAPbI3 with respect to the substrate curvature was substantially less pronounced.

The herein observed size reduction of microcrystallites concomitant with denser coverage of MAPbI3 deposits coated onto quartz cylindrical substrates with diminishing diameters was attributed to both the diminishing substrate heat capacities and, albeit to a lesser extent, the enhanced evaporation rates of the solvent for the substrates with increasing curvatures.

Finally, we also demonstrated that the cylinder-shaped geometry of the densely packed microcrystalline MAPbI3 deposit may serve as a model for simple, low-cost photooptical devices, including gas sensors.

**METHODS**

**Single-Step Solution Casting of MAPbI3 on CylinderShaped Substrates: Preparation of MAPbI3 Precursor in DMF Solution**. Prior to preparing the saturated DMF solutions of MAPbI3 it was important to prepare stoichiometric and pure single crystals. MAPbI3 single crystals were made by precipitation from an aqueous solution of concentrated hydroiodic acid (57 wt % in H2O) containing stoichiometric amounts of lead(II) acetate trihydrate (Pb(ac)2·3H2O) and methylamine (CH3NH2, 40 wt % in H2O). MAPbI3 crystals were grown and recrystallized in the saturated hydroiodic acid solution applying a temperature gradient in the vessel. The crystals were dissolved at the higher temperature side of the vessel and recrystallized at the lower temperature side of the vessel.44 Subsequently, the thus obtained MAPbI3 single crystals were harvested, dried at 120 °C, and dissolved in DMF, thus leading to a 50 wt % DMF stock solution.

**Single-Step Solution-Casting of MAPbI3 on CylinderShaped Substrates: Thin-Film Deposition**. All materials for fabrication of MAPbI3 deposits were purchased from SigmaAldrich and used as received. The thin polycrystalline layers of MAPbI3 were deposited via one-step solution casting on cylinder-shaped quartz substrates having the following diameters: 80, 330, 400, 700, 1100, and 1800 μm. Prior to the film deposition process, the quartz rods were carefully washed in acetone and ethanol, rinsed with the deionized water, and dried in air. Subsequently, the cylinder-shaped quartz substrates were covered with the stock solution of stoichiometrically mixed MAPbI3 precursors using dipping and doctor-blade techniques. Finally, the substrates were cured on a hot plate at 110 °C for 10 min. While curing, the rods were positioned at a small distance (3.0 mm) from the hot plate, thus allowing hot air to circulate all over the annealed substrate. The corresponding scheme of the experimental setup is shown in Figure S1. To ACS Photonics Article DOI: 10.1021/acsphotonics.7b01496 ACS Photonics 2018,

 avoid interaction with air and moisture, the prepared samples were stored under an inert atmosphere (nitrogen) in sealed glass containers.

**Scanning Electron Microscopy Imaging**. The morphology of the polycrystalline MAPbI3 layers deposited on cylindershaped quartz rods was investigated by SEM with the help of both secondary electrons and in-lens detectors. All SEM images were collected with a high-performance Schottky field-emission electron microscope capable of resolution in the 2−5 nm size range, model LEO 1550 (Carl Zeiss AG, Jena, Germany), in the Center of MicroNanoTechnology (CMi) at the EPFL. The analysis of the SEM images made it also possible to estimate the thickness of MAPbI3 films deposited on individual cylindershaped quartz substrates.

**Evaluation of the Surface Roughness of Quartz Substrates by Atomic Force Microscopy**. The surface roughness of uncoated quartz rod substrates was investigated by AFM. The visualization of the surface topography and the 2D roughness profiles were done with an AFM XE-100 from Park Systems Inc. (Santa Clara, CA, USA). The measurements were performed under contact mode using silicon cantilevers, model NSC36-10 M (spring constant of 5.0 N/m), from Park Systems. The analysis of variation of the substrate roughness was conducted with a standard Park Systems software package, XEI.

**Optical Microscopy Imaging and Steady-State Photoluminescence Measurements**. The optical microscopy imaging under visible light illumination, as well as the luminescence microscopy imaging and collection of the steady-state PL spectra, was performed using a customdesigned setup, which combined an inverted biological epifluorescence microscope (TC5500, Meiji Techno, Japan) with a compact spectrofluorometer (USB 2000+XR, Ocean Optics Inc., USA). A digital noncooled CCD camera (Infinity 2, Lumenera Co., Ottawa, Canada) was used to capture the luminescence images of the studied MAPbI3 deposits. The PL spectra and luminescence images were collected under illumination with an excitation wavelength $\lambda_{exc}$ = 546 nm. The excitation wavelength was filtered out from the emission spectrum of the microscope's Mercury vapor 100 W lamp by implementing the dedicated set of Meiji Techno filters, model 11002v2 green.[45]

Prior to performing luminescence microscopy imaging and collecting the PL spectra, the cylinder-shaped quartz substrates covered with MAPbI3 were centrally positioned in short sections of larger quartz capillaries (3.0 mm i.d./4 mm o.d.), which served as external, light-transparent sample holders. Centering the quartz rods inside the quartz tube was achieved by implementing two short sections of coiled copper wires (0.6 mm DIA), to which the terminations of the rods were glued. Afterward, both ends of the quartz tube (sample holder) were connected to the corresponding outlets of gaseous media, i.e., oxygen (O2), nitrogen (N2), and argon (Ar), using Luer-type plastic connectors and thin polypropylene flexible tubes. The whole setup was then attached to a microscope sample holder plate, as shown in Figure 4(a).

This approach made it possible to position the sample holder on the microscope XY stage and provided an easy way for the direct delivery of various gaseous media to the MAPbI3-covered substrate as well as enabled us to detect gas-dependent changes in the characteristic PL spectra of MAPbI3. Gas Flow. For our experiments we have used dry oxygen, nitrogen, and argon with a flow rate of ∼68 L/h. The flow rate was regulated and measured by a Swagelock variable area flowmeter (VAF-G2-06M-1-0).

**ACKNOWLEDGMENTS**

We would like to thank Center of MicroNanoTechnology (CMi) at EPFL for providing SEM facilities. This work was supported by ERC advanced grant "PICOPROP" (Grant No. 670918).

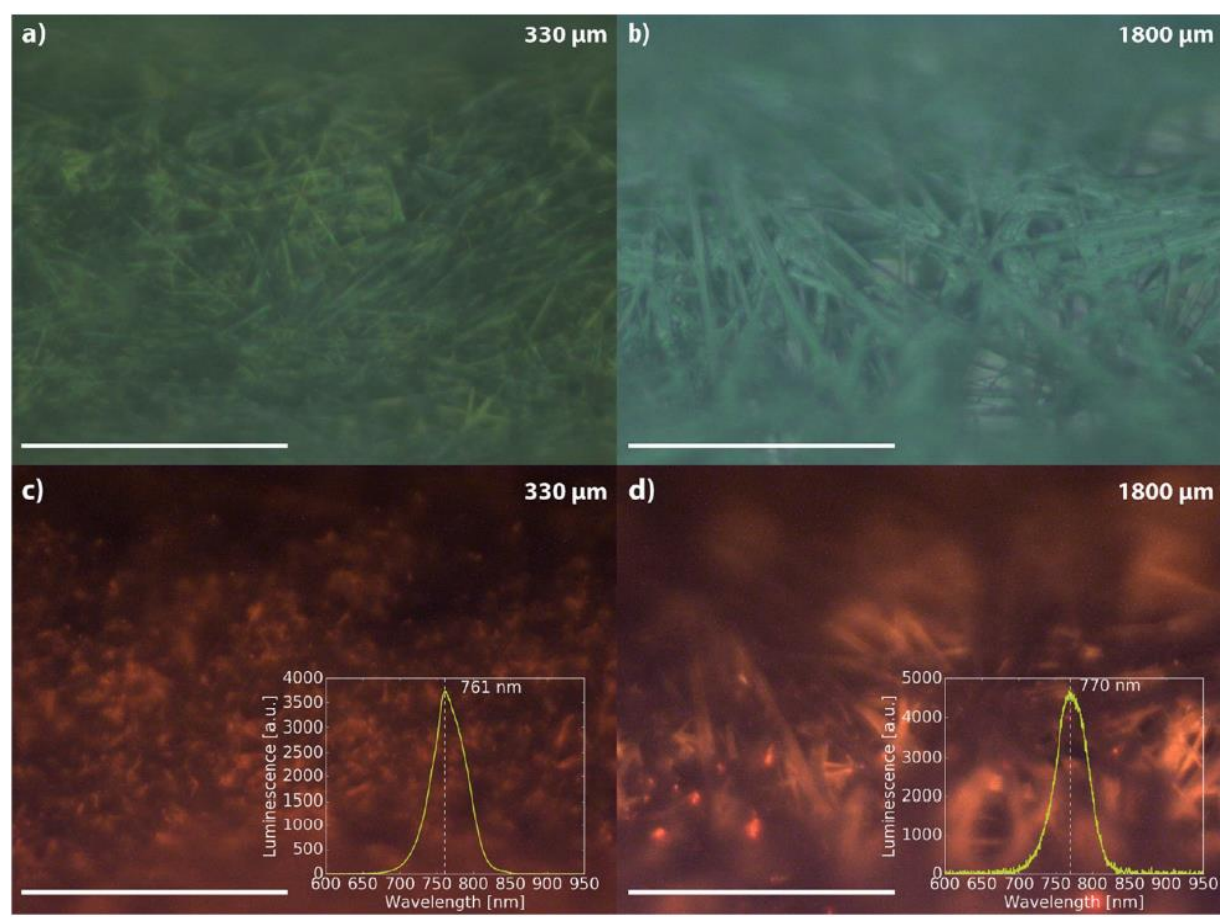

**Figure 1.** Comparison of the optical microscopy images of MAPbI3 deposits on cylinder-shaped quartz substrates having diameters of 330 and 1800 μm. Bright-field microscopic images acquired under visible light illumination for the cylinder-shaped substrates with diameters of 330 μm (a) and 1800 μm (b). Corresponding luminescence images collected under excitation with $\lambda_{exc}$ = 546 nm for the MAPbI3 films deposited on the substrates with diameters of 330 μm (c) and 1800 μm (d). Insets to (c) and (d): the steady-state PL spectra acquired simultaneously with luminescent imaging. In all images the scale bars are set to 50 μm.

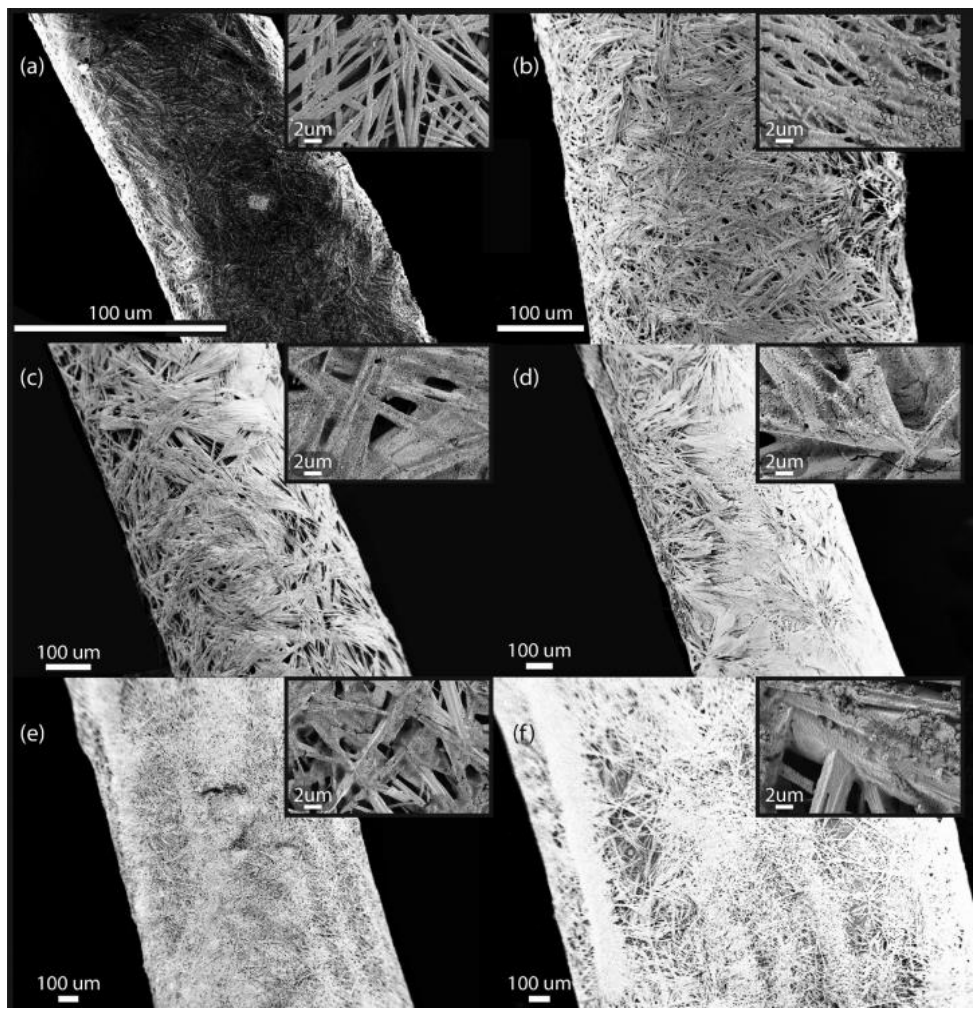

**Figure 2**. SEM images of microcrystalline MAPbI3 deposits on cylinder-shaped quartz substrates with varying diameters: (a) 80 µm, (b) 330 µm, (c) 400 µm, (d) 700 µm, (e) 1100 µm, and (f) 1800 µm. The insets show the magnified image of the deposits on the same size rods as the main figure.

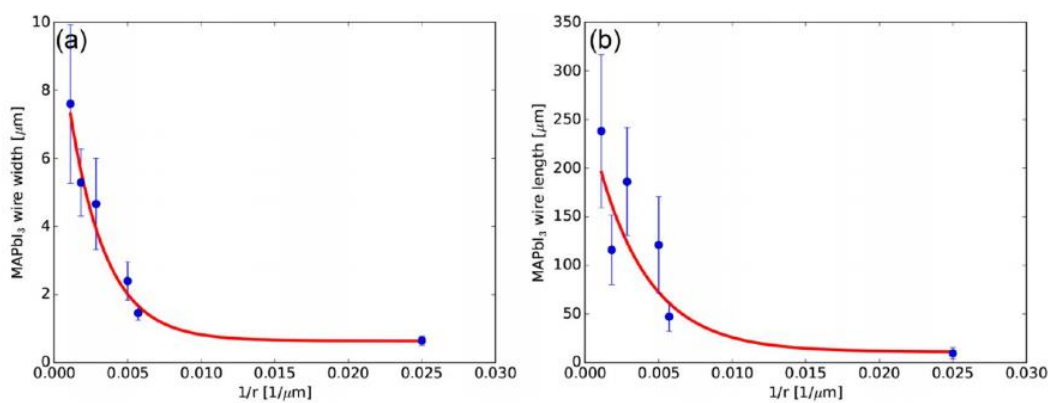

**Figure 3**. Average characteristic sizes of MAPbI3 microwires plotted as a function of the substrate curvature (1/r): (a) average widths and (b) average lengths. The solid line corresponds to the fit with an exponential function.

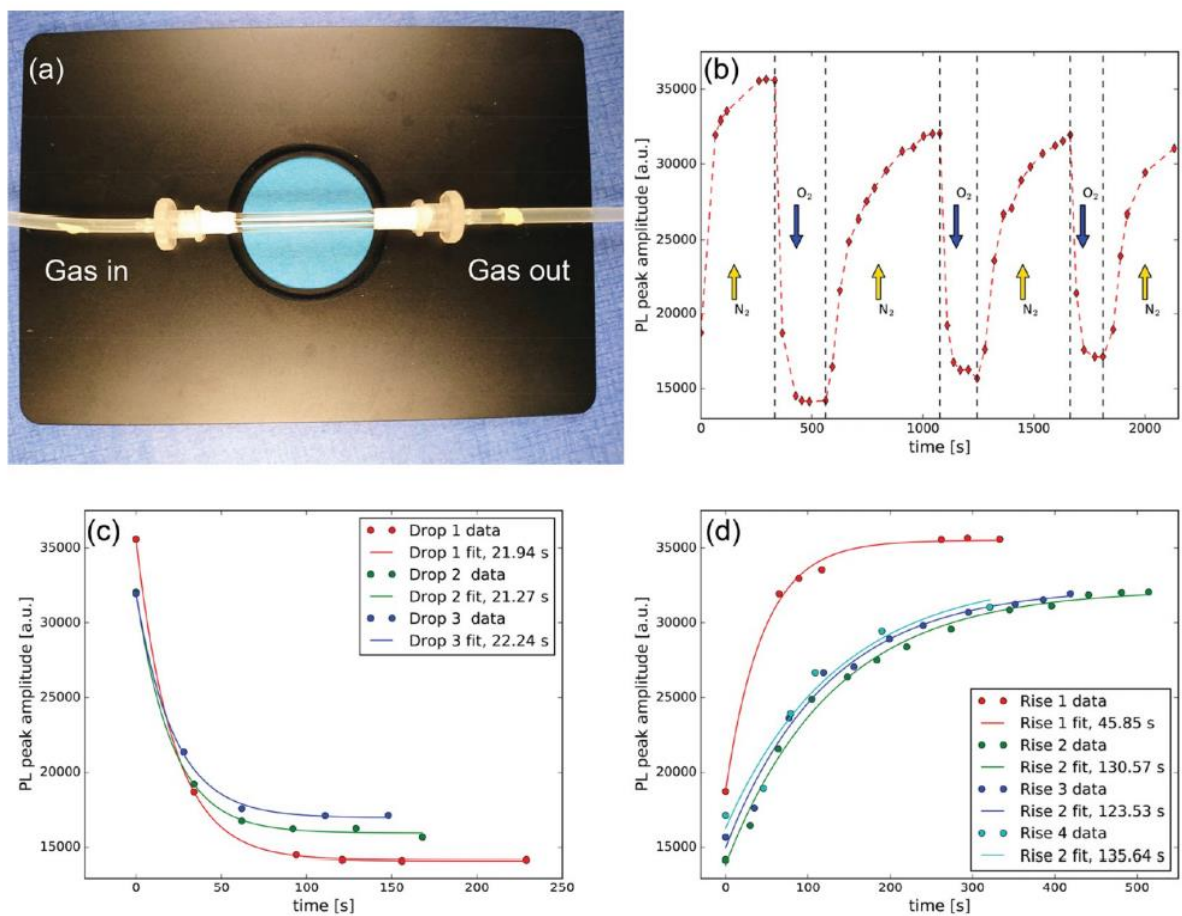

**Figure 4**. Experimental setup for acquiring PL spectra of MAPbI3 deposited onto cylinder-shaped substrates under exposure to the flow of various gaseous media and the example time-dependent evolution of the photoluminescent response of the light-soaked devices versus intermittent flow of N2 and O2. (a) Assembly consisting of a 330 μm quartz rod coated with MAPbI3, centrally positioned in a quartz tube (3.0 mm i.d./4.0 mm o.d.) and attached to a microscope sample holder plate. (b) Typical time evolution of the PL signal acquired for MAPbI3 deposited onto a cylinder-shaped substrate (quartz rod, 330 μm in diameter) under continuous excitation at λexc = 546 nm and intermittent exposure to N2 and O2 in the steady-state regime. (c) Overlapped plots of the decreasing portions of the photoluminescent response under exposure to O2. (d) Overlapped plots of the increasing portions of the photoluminescent response under exposure to N2. The solid lines correspond to the exponential fits. The steepest plot in (d) corresponds to the initial light-soaking treatment under exposure to the flow of N2 (red trace).